\documentclass[copyright,creativecommons]{eptcs}
\providecommand{\event}{TARK 2023} % Name of the event you are submitting to

\usepackage{iftex}

\ifpdf
  \usepackage{underscore}         % Only needed if you use pdflatex.
  \usepackage[T1]{fontenc}        % Recommended with pdflatex
\else
  \usepackage{breakurl}           % Not needed if you use pdflatex only.
\fi

    \def\draft{1}  % 0 to exclude author notes
    \def\ITCS{0} % 0 if not ITCS'20 version
    \ifnum\ITCS=0
    \fi
    \usepackage{verbatim}
    
    \ifnum\ITCS=1
\usepackage[utf8]{inputenc}
\usepackage{amsmath,amssymb,amsthm}
\usepackage{xcolor}
\usepackage[hidelinks]{hyperref}
\usepackage{bm}
\usepackage{nicefrac}
\usepackage{paralist}
\usepackage{typearea}
\typearea{16}
\fi

    \usepackage{tikz}
    \usetikzlibrary{positioning,chains,fit,shapes,calc}
    \usepackage{float,subcaption}
    \floatstyle{boxed}
    \restylefloat{figure, graphicx}

    \ifnum\ITCS=1
    \allowdisplaybreaks
\fi

    \usepackage{amsfonts,url,ifthen,epsfig,amsmath,amssymb,color,framed,float,mathrsfs}
    \usepackage{algorithm}
    \usepackage[noend]{algpseudocode}
    \usepackage{etoolbox}
    
    \makeatletter
    \let\original@footnotemark\footnotemark
    \newcommand{\align@footnotemark}{%
      \ifmeasuring@
        \chardef\@tempfn=\value{footnote}%
        \original@footnotemark
        \setcounter{footnote}{\@tempfn}%
      \else
        \iffirstchoice@
          \original@footnotemark
        \fi
      \fi}
    \pretocmd{\start@align}{\let\footnotemark\align@footnotemark}{}{}
    \makeatother
    
    \hypersetup{colorlinks=false}
    
    \ifnum\draft=1
    \newcommand{\Rnote}[1]{\begin{framed}\noindent \textcolor{red}{{#1}}\end{framed}} % boxed and red
    \else
    \newcommand{\Rnote}[1]{}
    \fi

    \newcommand{\remove}[1]{}
    \newtheorem{theorem}{Theorem}
    \newtheorem{example}{Example}

    \newtheorem{lemma}{Lemma}
    \newtheorem{corollary}{Corollary}
    \newtheorem{obs}{Observation}
    \newtheorem{definition}{Definition}
    \newtheorem{proposition}{Proposition}
    \newtheorem{remk}[theorem]{Remark}
    \newenvironment{remark}{\begin{remk} \begin{normalfont}}{\end{normalfont}
    \end{remk}}

    \def\FullBox{\hbox{\vrule width 8pt height 8pt depth 0pt}}
    
    \def\qed{\ifmmode\qquad\FullBox\else{\unskip\nobreak\hfil
    \penalty50\hskip1em\null\nobreak\hfil\FullBox
    \parfillskip=0pt\finalhyphendemerits=0\endgraf}\fi}
    
    \def\qedsketch{\ifmmode\Box\else{\unskip\nobreak\hfil
    \penalty50\hskip1em\null\nobreak\hfil$\Box$
    \parfillskip=0pt\finalhyphendemerits=0\endgraf}\fi}
    
 \ifnum\ITCS=0    
    \newenvironment{proof}{\begin{trivlist} \item {\bf Proof:~~}}
      {\qed\end{trivlist}}
      \fi

    \newenvironment{proofof}[1]{\begin{trivlist} \item {\bf Proof of #1:~~}}
      {\qed\end{trivlist}}

    \newcommand{\beq}{\begin{equation}}
    \newcommand{\eeq}{\end{equation}}
    \newcommand{\be}{\begin{enumerate}}
    \newcommand{\ee}{\end{enumerate}}
    \newcommand{\bi}{\begin{itemize}}
    \newcommand{\ei}{\end{itemize}}
    \newcommand{\bd}{\begin{description}}
    \newcommand{\ed}{\end{description}}
    \newcommand{\bc}{\begin{center}}
    \newcommand{\ec}{\end{center}}
    \newcommand{\bthm}{\begin{theorem}}
    \newcommand{\ethm}{\end{theorem}}
    \newcommand{\bdefi}{\begin{definition}}
    \newcommand{\edefi}{\end{definition}}
    \newcommand{\bcor}{\begin{corollary}}
    \newcommand{\ecor}{\end{corollary}}
    \newcommand{\blem}{\begin{lemma}}
    \newcommand{\elem}{\end{lemma}}
    \newcommand{\bexa}{\begin{example}}
    \newcommand{\eexa}{\end{example}}
    \newcommand{\bprop}{\begin{proposition}}
    \newcommand{\eprop}{\end{proposition}}

   \newcommand{\added}[1]{{\color{blue} #1}}
   \newcommand{\ul}{\underline}

    \ifnum\draft=1
    \newcommand{\ronote}[1]{\begin{framed}\noindent \textcolor{red}{{Ronen's note: #1}}\end{framed}} % boxed and red
    \newcommand{\ranote}[1]{\begin{framed}\noindent \textcolor{red}{{Rann's note: #1}}\end{framed}} % boxed and red
    \else
    \newcommand{\ranote}[1]{}
    \newcommand{\ronote}[1]{}
    \fi

    \def\real{\hbox{\rm\setbox1=\hbox{I}\copy1\kern-.45\wd1 R}}
    \def\neal{\hbox{\rm\setbox1=\hbox{I}\copy1\kern-.45\wd1 N}}

    \newcommand{\abs}[1]{\left|{#1}\right|}
    \newcommand{\eps}{\varepsilon}

    \newcommand{\scM}{\mathcal{M}}

    \newcommand{\R}{{\mathbb R}}

    \definecolor{myblue}{RGB}{80,80,160}
    \definecolor{mygreen}{RGB}{80,160,80}
    
    \title{Selling Data to a Competitor\\ \footnotesize{(Extended Abstract)}}
 \ifnum\ITCS=1
 \author{}
 \else   
    \author{Ronen Gradwohl \institute{Department of Economics and Business Administration\\ Ariel University} \email{roneng@ariel.ac.il}
    \and
    Moshe Tennenholtz \institute{Faculty of Data and Decision Sciences\\ The Technion -- Israel Institute of
    Technology} \email{moshet@ie.technion.ac.il}
    }
    \def\titlerunning{Selling Data to a Competitor}
\def\authorrunning{R. Gradwohl \& M. Tennenholtz}

    \fi
\begin{document}

	\maketitle

       \begin{abstract}
       
We study the costs and benefits of selling data to a competitor. Although selling all consumers' data may decrease total firm profits, there exist other selling mechanisms---in which only some consumers' data is sold---that render both firms better off. We identify the profit-maximizing mechanism, and show that the benefit to firms comes at a cost to consumers. We then construct Pareto-improving mechanisms, in which each consumers' welfare, as well as both firms' profits, increase. Finally, we show that consumer opt-in can serve as an instrument to induce firms to choose a Pareto-improving mechanism over a profit-maximizing one.

% We study the effects of data sharing between firms on prices, profits, and consumer welfare. Although indiscriminate sharing of consumer data decreases firm profits due to the subsequent increase in competition, selective sharing can be beneficial. We show that there are data-sharing mechanisms that are strictly Pareto-improving, simultaneously increasing firm profits and consumer welfare. Within the class of Pareto-improving mechanisms, we identify one that maximizes firm profits and one that maximizes consumer welfare.

%       
       \end{abstract}

%    \noindent \textbf{JEL Classification:} C72, D82\\
%    \noindent \textbf{Keywords:} 
%    \thispagestyle{empty}
%    
%\end{titlepage}
%    
    
    \renewcommand{\thefootnote}{\arabic{footnote}}
    \setcounter{footnote}{0}

\section{Introduction}

In recent years, it has become common wisdom that data is a dominant source of power. This power is perhaps most clearly illustrated in markets where an incumbent with access to consumer data competes with an entrant who does not have such data. As stated in a crisp manner by \cite{macmillan2008incumbent}, common wisdom holds that  the incumbent's key advantage is data superiority: ``If you run a market-leading company, you should never be blindsided by an invader. Locked within your own records is a huge, largely untapped asset that no attacker can hope to match: what we call the incumbent's advantage.'' The situation is not uncommon: In our data-driven economy, competing firms often find themselves in asymmetric situations where one of them has superior or even exclusive access to relevant data.

Such data asymmetry has  become a major issue for debate. For example, in a June 2021 press release, the EU declared that it has opened an antitrust investigation that will ``examine whether Google is distorting competition by restricting access by third parties to user data for advertising purposes on websites and apps, while reserving such data for its own use,'' \cite{EU2021}. 
One of the issues in such debates is the question of data sharing: Should the incumbent share its data in order to increase market competition and consumer welfare? And can the incumbent profitably sell its data to the entrant?

%``has opened a formal antitrust investigation to assess whether Google has violated EU competition rules by favouring its own online display advertising technology services in the so called `ad tech' supply chain, to the detriment of competing providers of advertising technology services, advertisers and online publishers. The formal investigation will notably examine whether Google is distorting competition by restricting access by third parties to user data for advertising purposes on websites and apps, while reserving such data for its own use'' \cite{EU2021}.
%Without digging into the above anti-trust issues, the above points to a major issue we face in our data economy: competing firms might find themselves in an asymmetric situation where data about customers is held by one of the parties;  one of them may have superior or even exclusive access to relevant data.

Strategic decisions about the sale of data to competitors appear in the online economy frequently, although they are not always stated explicitly. For example, when an advertiser buys a sponsored-search campaign through an ad exchange, the advertiser obtains useful information about a segment of consumers as part of the ad exchange reports. The advertiser might then use this information when bidding directly for display ads on other platforms, including platforms on which the ad exchange is also a competitor. The data-holder (e.g., ad exchange) thus faces a strategic decision about which consumer segments to sell to a competitor, and at what prices. The data-buyer (e.g., advertiser), in turn, must decide whether to pay the price and obtain data about these consumer segments, or whether to enter into competition without the data on offer.

%More generally, tools that facilitate data-sharing between an incumbent and In the online advertising market, for example, firms use their data
%to personalize advertisements in order to maximize the probability that a consumer clicks on their ad. 
%Data sharing is common in this market, and platforms such as Google Merchant and Azure Data Share 
%specifically facilitate such sharing across clients in order to improve personalization and increase clickthrough rates.

These strategic considerations raise numerous questions: What are the data-holder's costs  and benefits from selling data to a data-buyer? What are the effects of data sale on consumer welfare? And, in a regulated market, can data sale be regulated in a way that leads to Pareto-improving transactions, benefitting consumers as well as firms?
   
In this paper we study these questions in a paradigmatic model of imperfect competition between two firms who have asymmetric access to data.
We consider the classic Hotelling model of imperfect competition: There are two firms, each located at a different endpoint of a unit interval, with a unit mass of consumers distributed across this interval. We model data about a consumer as information about the consumer's location on the interval. In the classic model, neither firm has data about any consumer, and so firms engage in competition via uniform prices that each offers to all consumers. In our variation of this model, in contrast, one firm is a data-holder who knows the locations of all consumers, whereas the other firm is a data-buyer who has no such data. The data-holder can use its data advantage in order to personalize prices to consumers, and can thus sometimes undercut the data-buyer's uniform price.

In order to study the costs and benefits of data sale to a competitor, we suppose the firms engage in a data-sharing mechanism. Such a mechanism consists of a segment of consumers whose data the data-holder shares with the data-buyer, as well as a price the data-buyer pays the data-holder. After engaging in such a mechanism, the data-buyer will hold location data about all consumers in the shared segment, allowing that firm to also personalize prices to them.

Within this model, we first show that full data-sharing, in which the data-holder shares all its data with the data-buyer, is harmful to the firms. We then show that there exist other data-sharing mechanisms---in which only some consumers' data is shared---that increase both firms' profits. In fact, we identify the mechanism that maximizes total firm profits. This last mechanism, however, increases firm profits at the expense of consumers. We thus proceed to show that there exist Pareto-improving mechanisms, in which each consumers' welfare, as well as both firms' profits, increase. Finally, we consider the question of how a regulator can induce firms to utilize a Pareto-improving mechanism rather than a profit-maximizing one that may harm consumers. We show that consumer opt-in may serve as such an instrument: If consumers are given the opportunity to opt-in to having their data sold, and if the data-holder is only permitted to share data about consumers who have opted-in, then in equilibrium the firms will choose a Pareto-improving mechanism.
    
Our results are driven by two forces, which we identify as the direct effect and the indirect effect of data sharing. The direct effect is the following: if the data-holder shares data about a particular consumer, then the data-buyer can now offer that consumer a personalized rather than the uniform price. This affects both firms' equilibrium personalized prices to that consumer, and may thus impact profits and welfare. The indirect effect of data sharing, on the other hand, is the following: by sharing data about a segment of consumers, the data-holder changes the set of consumers to whom the data-buyer's uniform price applies (since additional consumers will now be offered personalized
prices). And since the uniform price is determined in equilibrium in part by the locations of consumers to whom that price will
apply, a change in the set of consumers may effect a change in the equilibrium uniform price, thereby affecting profits and welfare. Our results highlight how the interplay between the direct and indirect effects of data sharing lead to changes in firms' profits and consumers' welfare.

In addition to identifying the two effects of data sharing, our analysis generates several general insights. First, and perhaps surprisingly, selling data to a competitor can be strictly beneficial to both firms.\footnote{We note that this holds even if data is sold at no cost---see Proposition~\ref{prop:firm-optimal}.} Second, data can be sold in a way that is Pareto improving. And finally, such Pareto-improving data-sale can be induced by consumer opt-in regulation.

We note that the idea of selling data to a competitor has been advocated in financial markets (see, for example, \cite{admati1988selling}). In that context, the possibility of data sale allows a decision maker to choose between taking investment risks or obtaining direct monetary rewards. The incredibly fast-growing data-economy, in which some firms hold massive amounts of data about consumers, raises calls to consider such data sale in a broader context: Can it lead to increased profits to both data-holders and data-buyers? And can it benefit all of society, including consumers whose data is exchanged?

%
%\paragraph{Organization of the paper} Immediately following is a review of the related literature, after which we formally describe the model. The subsequent sections then contain our various analyses: Section~\ref{sec:no-sharing} on the baseline case of no data-sharing, Section~\ref{sec:full-sharing} on full data-sharing, Section~\ref{sec:firm-optimal} on firm-optimal data-sharing, and Section~\ref{sec:pareto-improving} on Pareto-improving data-sharing. Finally, in Section~\ref{sec:opt-in} we present our results on consumer opt-in, after which we conclude.

%Our analysis then proceeds in several stages of increasing generality. We begin in Section~\ref{sec:model} with an analysis of data sharing in the simple case where only one firm has data about consumers, whereas the other has no data at all. Although interesting in their own right, the mechanisms we design here also help to develop intuition and serve as building blocks for our subsequent mechanisms in Sections~\ref{sec:two-segments} and~\ref{sec:four-segments}, in which both firms are assumed to have some data about consumers. Throughout, we make a standard distributional assumption for Hotelling games, namely, that consumers are uniformly distributed within the unit interval. As a robustness check, in Appendix~\ref{sec:non-uniform} we relax this assumption and generalize our main result---the construction of a firm-optimal, Pareto-improving mechanism---to general consumer distributions.

\smallskip
\textbf{Related literature}~~~
%Competition in ML and prediction: \cite{mansour2018competing}, \cite{ben2019regression}, \cite{feng2019bias}, \cite{gradwohl2020coopetition}
%
This paper is part of a large and growing literature on data markets (see, e.g., the survey of \cite{bergemann2019markets}). Work in this area focuses on related but orthogonal questions, such as the effects of data-sale by a third-party data-provider and of information sharing between competitors. Our paper bridges these strands by considering data sale to a competitor. To the best of our knowledge, the work of \cite{admati1988selling} is the only other paper to study such a scenario, and ours is the first to focus on the effects of such data sale on firm profits as well as consumer welfare.

The literature on the sale of data by a data provider (e.g., \cite{admati1986monopolistic,bergemann2018design,montes2019value,segura2021selling,yang2022selling} and others) studies how a third-party data-provider can maximize profits by selling data to a monopolist or to competing firms who use this data to price discriminate. 
Within this literature, one paper that is closely related to ours is that of \cite{elliott2021market}. \cite{elliott2021market}  consider an information designer who provides consumer information to oligopolists, and characterize the different market outcomes that can be achieved by the designer. Our paper differs from this research in that we suppose data is not held by a third-party, but rather by one of the competing firms. This firm may sell data to its direct competitor,  affecting both firms'  respective market positions. 

A different but related setting is that of \cite{ali2020voluntary}, where the consumers are holders of information who may share it with one or both firms so as to intensify competition.  The model and results of \cite{ali2020voluntary}  bear some similarity to ours. For example, they also consider a Hotelling model, and show that consumers are better off whenever those sufficiently closer to one firm than another share their location with that firm, and those closer to the middle share their location with both firms. Despite the similarities, our paper studies an orthogonal question, as we assume one of the firms already has data about consumers, and focus on whether that firm will sell data to its competitor. In contrast to the model of \cite{ali2020voluntary}, in which each consumer chooses which firm has access to that consumer's location, in our model the informed firm strategically chooses whether or not to share this information. Under consumer opt-in the role of consumers is in determining whether such sharing could potentially take place, but not in whether it actually takes place. Finally, while \cite{ali2020voluntary} show that, consumers are always better off when they share some of their information, we show that when firms choose what information to share this may no longer be the case.
%: even under consumer opt-in, there exist equilibria in which consumers are worse off.

Because our work considers the sale of data from one firm to another, it is related to the literature on information sharing.
Although information sharing between firms has been studied in a variety of settings,\footnote{These include oligopolistic competition \cite{clarke1983collusion,raith1996general}, financial intermediation \cite{pagano1993information,jappelli2002information,gehrig2007information}, supply chain management \cite{ha2008contracting,shamir2016public},
competition between data brokers \cite{gu2022data,ichihashi2020competing}, and advertising \cite{gradwohl2023coopetition}.} our paper is most-closely related to that of competitive price discrimination---see, for example, the surveys of \cite{stole2007price} and \cite{fudenberg2012digital}. One of the main insights from this literature is that when firms have more data about consumers, competition between them is more intense, leading to lower prices and profits. 
% And although this is generally beneficial to consumers, it harms firms. An immediate corollary is that, in general, full data-sharing (which leads to firms having more data about consumers) is harmful to firms, echoing the concern quoted above from the \cite{EC2020}.
In our paper, in contrast, data is sold by one firm to another  in such a way as to increase profits.

%Papers that specifically analyze the effects of data sharing within a Hotelling model include %\cite{chen2001individual}, \cite{shy2013investment}, 
Two papers that specifically analyze the effects of data sharing within a Hotelling model are \cite{jentzsch2013targeted} and \cite{braulin2023effects}.
\cite{jentzsch2013targeted} study a model in which each of two firms may have data both about consumers' locations and about their transportation costs, and consider the eight permutations in which each firm may have either a dataset about locations, a dataset about transportation costs, both datasets, or neither datasets. They then analyze the market effects of firms sharing one or both of their (full) datasets with each other, and provide conditions under which sharing is beneficial to the firms. 
%In particular, it is harmful for firms to share both datasets, but may be beneficial for them to share their (full) data on transportation costs alone. They also show that such partial sharing is typically detrimental to consumers.
\cite{braulin2023effects} studies a Hotelling model in which locations are two-dimensional, and firms hold all data about one dimension, both dimensions, or neither dimension. He analyzes the various scenarios in terms of firm profits and consumer welfare, with a particular emphasis on the comparison to the regimes of full privacy (neither firm has any data) and no privacy (both firms have full data). Interestingly, \cite{braulin2023effects} shows that total firm profits are hump-shaped in the amount of information they hold; for example, the scenario in which each firm holds data about a different dimension yields higher profits than both full privacy and no privacy. 
%Unlike our paper, \cite{clavora2021effects} does not focus on the effects of data sharing, nor on the possibility of sharing partial data about a particular dimension. In addition, in his setup, changing the informational allocation typically has an ambiguous effect on consumers. In contrast, we focus on the possibility of designing data-sharing mechanisms in a way that increases all participants' welfare.
Our work differs from both of these papers in that we study the sale of partial data from one firm to another, with an emphasis on mutually increasing profits.

In terms of modeling, our paper is most closely related to \cite{montes2019value} and \cite{gradwohl2022pareto}.  \cite{montes2019value} consider a one-dimensional Hotelling model in which consumers' locations may be known to one, both, or neither firm. Their concern is not the sale of data from one firm to another, but rather the optimal strategy of a data broker who sells the data to the firms. They also consider the effects of a consumer-side technology that allows consumers the ability to protect their privacy. \cite{gradwohl2022pareto} also study a Hotelling model, but suppose that both firms have some data about consumers. Their main focus is on various forms of mutual data sharing between the firms.

\section{The Model}\label{sec:model}
%\subsection{Model}

We focus on a standard Hotelling model, in which a unit mass of consumers is spread over the unit interval according to an atomless distribution $F$ with continuous, strictly positive density $f$ that has full support.
There are two firms: firm $A$ is located at $\theta_A=0$, and
firm $B$ is located at $\theta_B=1$. Each consumer chooses at most one firm from which to purchase a good.
Consumers derive value $v$ from the good, but pay two costs: the price, and a linear transportation cost that scales
with the distance between the consumer and the firm providing the good.
Thus, a consumer located at $\theta$ who buys from firm $i$ at price $p_i$ obtains utility $v-p_i-t\abs{\theta-\theta_i}$, where
$t$ is the marginal transportation cost. 
We assume throughout that the market is covered---namely, that $v> 2t$---so that all consumers
purchase a good even when there is a monopolist firm. Finally, we also assume for simplicity that firms' marginal costs are 0, and
so their profit from the sale of a good is equal to the price. These are all standard assumptions in Hotelling games.

The standard setup consists of a two-stage game: First, firms simultaneously set prices; second, consumers choose a firm
and make a purchase. In the simple case where the distribution $F$ of consumers is uniform the game has a unique subgame perfect equilibrium: firms' prices are $p_A=p_B=t$, consumers
in $[0,0.5)$ buy from $A$, and consumers in $(0.5,1]$ buy from $B$ (see, e.g., \cite{belleflamme2015industrial}).\footnote{The equilibrium is unique up to the choice of the indifferent consumer located at $\theta=0.5$.}

In this paper we will consider a variant of the standard model by supposing that firms may have additional information about
some of the consumers. In particular, we will suppose that, for each consumer, one or both firms know the
location of that consumer on the unit interval. For such consumers, firms will be able to offer a {\em personalized price}---a 
special offer specifically tailored to that consumer. If a firm does not know a consumer's location,
however, then it cannot distinguish between that consumer and all other consumers whose location it does not know. All such consumers are offered the same {\em uniform price}.
%The firm will have the opportunity to choose a {\em uniform price}---a price it charges to all such indistinguishable 
%consumers---and a personalized price---a price it tailors for each consumer whose location it does know.

In our model, firm $B$ is the data-holder and firm $A$ is the data-buyer. Thus, initially, we assume that firm $B$ knows the locations of all consumers, whereas firm $A$ does not know any consumer's location. 
Given this informational environment, a data-sharing mechanism $M=(M_B,r)$ between firms specifies a subset $M_B\subseteq[0,1]$ and a number $r\in\R$, with the interpretation that firm $B$ shares with
firm $A$ the locations of consumers in $M_B$, and firm $A$ transfers to firm $B$ a payment $r$.
%\footnote{In Section~\ref{sec:no-transfers} we restrict attention to mechanisms with no transfers, where $r=0$.}
Two simple examples of data-sharing mechanisms are one that involves {\em no sharing}, $M=(\emptyset,r)$, and one that involves {\em full sharing}, $M=([0,1],r)$.
Alternatively, firm $B$ may share data about a subset of  consumers.  For example, under mechanism $([x,y],r)$, if consumer $\theta\in [x,y]$ arrives, both firms will know that consumer's location. On the other hand, 
if consumer $\theta\in [0,1]\setminus [x,y]$ arrives, firm $B$ will know that consumer's location, and firm $A$ will only be able to deduce that the consumer is not located within $[x,y]$.

In our analysis, we consider the following order of events:
\begin{enumerate}
\item Firms engage in a data-sharing mechanism $M=(M_B,r)$.
\item Firm $A$ announces uniform price $p_A$.\footnote{Note that firm $B$ knows all consumers' locations, and so personalizes prices to each. It therefore need not post a uniform price.}
\item A consumer arrives, and all firms who know the consumer's location $\theta$ simultaneously offer 
that consumer a personalized price, $p_A(\theta)$ and $p_B(\theta)$.
\item The consumer chooses a firm from which to buy, and payoffs are realized.
\end{enumerate}

Note that firms share data, and firm $A$ announces its uniform prices, before consumers arrive. 
After a consumer arrives to the market, the firms who know
the consumer's specific location simultaneously offer personalized prices. 
If firm $A$ offers a consumer a personalized price, this offer subsumes the firm's original uniform price.
Thus, the uniform price $p_A$ will apply only to those consumers who will not subsequently be offered a personalized price
by firm $A$.

Importantly, when firms set personalized prices, they know
the uniform price set by firm $A$ in the previous stage. 
This is the standard timing considered in the literature (see, e.g., \cite{thisse1988strategic,choudhary2005personalized,choe2018pricing,montes2019value,chen2020competitive}).\footnote{An alternative model that we do not analyze
is one in which firms set uniform and personalized prices simultaneously, for each consumer. \cite{montes2019value} show that, in this case, a (pure) equilibrium may fail to exist.}
%Also, \cite{fudenberg2000customer,liu2006customer} have first-period uniform prices followed by second-period personalized prices.
%Before data sharing, each firm knows the location of some subset of the consumers.

For any fixed mechanism $M$, we will consider the pure subgame perfect equilibria of the game that starts with data-sharing mechanism $M$. Such equilibria always exists, and consist of a uniform price for firm $A$ followed by personalized prices for both firms. Once the uniform price is fixed, the equilibrium personalized prices for each consumer $\theta$ are uniquely fixed. We will be interested in designing mechanisms $M$ that lead to equilibria with high firm-profits and high consumer-welfare.

One important desideratum of data-sharing mechanisms (with corresponding equilibria) is that they be {\em individually rational (IR)}: That the expected utility of each firm with data sharing be at least as high as without data sharing. A data-sharing mechanism should be IR if we expect firms to participate.

Our main focus will be on mechanisms that are not only IR, but also {\em Pareto-improving}:  when  sharing takes place, (i) the expected utility of each firm and {\em every} consumer be at least as high as without data sharing, and that (ii) either firm $A$'s profits, firm $B$'s profits, or total consumer welfare be strictly higher.

We note that many of our results make no assumptions about the distribution of consumers. In such a general setting there may be multiple equilibria, even with no data-sharing, each with different uniform prices. Hence, we will often describe mechanisms as being IR or Pareto-improving {\em relative to} a particular no-sharing equilibrium.

\section{No Data-Sharing}\label{sec:no-sharing}
We begin by analyzing equilibria under no data-sharing. To this end,
define
$\mu(p_A) = \frac{1}{2}-\frac{p_A}{2t}$. If firm $A$ charges uniform price $p_A$, then the consumer located at $\mu(p_A)$ is indifferent between purchasing from firm $A$ at that price and purchasing from firm $B$ at price 0. All consumers located to the left of $\mu(p_A)$ will thus strictly prefer purchasing from firm $A$ at price $p_A$ than from firm $B$ at any nonnegative price. In contrast, for every consumer located to the right of $\mu(p_A)$ there exists a nonnegative price of firm $B$ such that that consumer will prefer to purchase from $B$ than from $A$.
This is formalized in the following proposition:
\begin{proposition}\label{prop:no-sharing}
Let $P_A = \arg\max_{p}p\cdot F\left(\mu(p_A)\right)$.
Without data sharing, the set of equilibria consist of any uniform price $p_A\in P_A$ for firm $A$ and corresponding personalized prices  $p_B(\theta)=\max\{0,  p_A+t(2\theta-1)\}$ for firm $B$. In the equilibrium with uniform price $p_A\in P_A$, consumers in $[0, \mu(p_A))$ purchase from firm $A$, whereas consumers in $[\mu(p_A), 1]$ purchase from $B$. The equilibrium with $p_A=\max \{P_A\}$ is strictly dominant for the firms.
\end{proposition}
The proof of Proposition~\ref{prop:no-sharing}, and all other propositions, appear in the full version of this paper \cite{gradwohl2023selling}.

Throughout the paper we will illustrate our results with the simple case in which consumers are uniformly distributed on $[0,1]$. We note that this is the standard setup in Hotelling games.

\begin{example}\label{ex:uniform-no-sharing}
When consumers are uniformly distributed on $[0,1]$, the set $P_A=\{t/2\}$. In the unique equilibrium, then, consumers between 0 and $\mu(t/2)=1/4$ purchase from $A$ at uniform price $p_A=t/2$, whereas the rest purchase from $B$ at personalized prices $p_B(\theta)=\max\{t(2\theta-1/2),0\}$. Total firm profits are $\pi_A=t/8$ and 
$$\pi_B = \int_{1/4}^1 t(2\theta-1/2)d\theta = \frac{9t}{16},$$
whereas consumer welfare is
$$CW = \int_0^1 \max\{v-\theta t - p_A, v-t(1-\theta) - p_B(\theta)\} d\theta 
=  \int_0^1 (v - t/2 - \theta t)  d\theta = v-t.$$
\end{example}

%How does data sharing between the firms impact profits and welfare? This is the question we now proceed to answer.

%\Rnote{When have arbitrary $F$, there may be multiple equilibria (since firm $A$'s maximization may have several optima). $A$ is indifferent between all of them. If we choose the {\em highest} such $p_A$, then this the strictly optimal one for $B$. Result is basically about Pareto improvement relative to firm-optimal no-sharing equilibrium. In uniform case, the equilibrium is unique. (In contrast with \cite{ali2020voluntary}, where result is that lowest $p_A$ is optimal for consumers.)}

\section{The Direct Effect and Full Data-Sharing}\label{sec:full-sharing}
In this section we begin our analysis of how data-sharing impacts profits and welfare. Data sharing has a direct effect and an indirect effect. The direct effect is that if firm $A$
obtains information about a consumer's locations via the sharing mechanism, it can now offer that consumer a personalized price. This affects firm $B$'s equilibrium personalized price to that consumer, and hence also profits and welfare. The indirect effect of data sharing is that it may change
the set of consumers to whom firm $A$'s uniform price applies, since additional consumers will now be offered personalized
prices. And since the uniform price is determined in equilibrium in part by the locations of consumers to whom that price will
apply, a change in the set of consumers may effect a change in the equilibrium uniform price. 
In this section we explore the direct effect, and then in Section~\ref{sec:firm-optimal} we explore the indirect effect.

Suppose that, absent data-sharing, firm $A$'s uniform price is $p_A$.
If firm $B$ shares the location $\theta$ of some consumer with $A$, then the firms compete in personalized prices over that consumer, yielding equilibrium prices $p_A(\theta) = \max\{t(1-2\theta),0\}$ and $p_B(\theta)=\max\{t(2\theta-1),0\}$.
The direct effect of firm $B$ sharing the location of a consumer  is summarized in Lemma~\ref{lem:direct-effects}:
\begin{lemma}\label{lem:direct-effects}
Consider mechanism $M=(\{\theta\},0)$ relative to no sharing, and suppose that consumer $\theta$ shows up.
 \begin{enumerate}
\item If $\theta\in(1/2, 1]$, consumer $\theta$ still buys from $B$, but now at price $t(2\theta-1)$. This is a net loss of $p_A$ to firm $B$ and a net gain of $p_A$ to the consumer.
\item If $\theta\in[\mu(p_A),1/2)$,  consumer $\theta$ switches to purchasing from $A$, at price $t(1-2\theta)$.
This is a loss of $p_A+t(2\theta-1)$ to firm $B$, a gain of $t(1-2\theta)$ to firm $A$, and a gain of 
$p_A-t(1-2\theta)\geq 0$
to the consumer. Also, the gain to $A$ is greater than the loss to $B$ if and only if 
$$\theta<\frac{1}{2}\left(\mu(p_A)+\frac{1}{2}\right),$$
the midpoint of the interval of $\theta$-s in the case under consideration.
\item If $\theta\in[0,\mu(p_A))$, consumer $\theta$ still buys from $A$, now at personalized price $p_A(\theta)=t(1-2\theta)>p_A$.
\end{enumerate}
\end{lemma}

%\begin{proof}
%We prove each bullet in order:
%\begin{enumerate}
%\item If $\theta\in(1/2, 1]$, then, by Proposition~\ref{sec:no-sharing}, under no sharing consumer $\theta$ buys from $B$ at price $p_A+t(2\theta-1)$. Under $M$, both firms personalize prices to $\theta$, but because $\theta$ is closer to $B$ that firm will be able to charge a lower price. The maximal price firm $B$ can charge and still sell to $\theta$ is $p_B(\theta)=t(2\theta-1)$, since firm $A$'s personalized price to $\theta$ is 0.
%
%\item 
%If $\theta\in[\mu(p_A),1/2)$ and $p_A(\theta)=t(1-2\theta)$, consumer $\theta$ will prefer to purchase from $A$ rather than from $B$ even when $p_B(\theta)=0$.
%The consumer's gain is $$p_A+t(2\theta-1)+t(1-\theta) - t(1-2\theta) - t\theta = p_A-t(1-2\theta)\geq 0,$$
%where the inequality follows from $\theta\geq\mu(p_A)$.
%The second inequality follows since
%$$\theta<\frac{1}{2}-\frac{p_A}{4t} = \frac{\mu(p_A)}{2}+\frac{1}{4}=\frac{1}{2}\left(\mu(p_A)+\frac{1}{2}\right).$$
%
%\item If  $\theta\in[0,\mu(p_A))$ and $p_A(\theta)=t(1-2\theta)$, consumer $\theta$ will still prefer to purchase from $A$. The inequality $p_A(\theta)=t(1-2\theta)>p_A$ follows from $\theta<\mu(p_A)$.
%\end{enumerate}
%\end{proof}

Given these direct effects, we now consider full data-sharing, namely, $M=([0,1],r)$ for some $r$.
Under this mechanism, both firms know the location of every consumer, and so both engage in personalized
pricing. Firm $A$'s uniform price thus applies to no consumer, and so only the direct effect has any bite.
By Lemma~\ref{lem:direct-effects}, relative to the no-sharing mechanism with price $p_A$, consumers  $\theta\in[\mu(p_A),1]$ are better off, whereas consumers $\theta\in[0,\mu(p_A))$ are worse off, under full data-sharing. For the firms, naturally firm $B$ is better off with no sharing and firm $A$ with full sharing. 
The effect on total profits, however, depends on the distribution $F$. 
For the case of uniformly distributed consumers, full data-sharing harms firms:

\begin{example}
When consumers are uniformly distributed, \cite{taylor2014consumer} show that profits are $\pi_A=\pi_B=t/4$ (see also \cite{thisse1988strategic}).
% and consumer welfare is $CW = v-3t/4$.
Note that total profits $\pi_A+\pi_B$ are higher under no sharing ($t/8 + 9t/16 = 11t/16$, by Example~\ref{ex:uniform-no-sharing}) than under full sharing ($t/4+t/4=t/2$). 
This implies that no mechanism $([0,1],r)$ is IR, regardless of $r$.
\end{example}
Although full data-sharing decreases total firm profits when consumers are uniformly distributed, there exist distributions of consumers under which full data-sharing increases profits---for example, this is the case when a $(1-\eps)$-fraction of consumers are uniformly distributed on the sub-interval $[0,1/4]$, and the remaining $\eps$ on $(1/4,1]$, for some small enough $\eps>0$.\footnote{Such a consumer distribution does not satisfy our continuity assumption. However, the same result holds also when the kink at $1/4$ is smoothed out.} 
However, even then full sharing does not lead to {\em maximal} profits. We now turn to mechanisms that do.

\section{The Indirect Effect and Firm-Optimal Data-Sharing}\label{sec:firm-optimal}
In this section we describe firm-optimal mechanisms, which exploit the {\em indirect} effect of data sharing. By Lemma~\ref{lem:direct-effects}, firm $B$'s profit from a consumer $\theta\in(1/2,1]$ is $p_A+t(2\theta-1)$. If $A$'s uniform price were to increase, this would likewise increase $B$'s profit from consumer $\theta$. Now, recall that, when there is no sharing, firm $A$ sets its uniform price by choosing $p_A\in P_A = \arg\max_{p}p\cdot F\left(\mu(p)\right)$. If $B$ were to share data about consumers in some interval $[\ul\theta, \mu(p_A)]$, however, then $A$ would offer consumers on this interval a personalized price. The uniform price would no longer apply to them, but would instead apply only to consumers $[0,\ul\theta)\cup(\mu(p_A),1]$. Firm $A$ may then benefit from increasing (decreasing) the uniform price above (below) $p_A$, at the same time increasing (decreasing) the profits of firm $B$ from consumers $\theta\in(1/2,1]$. This is the indirect effect of data sharing.

%Given these two effects, there are two reasons firm $B$ might benefit from sharing the location of a consumer $\theta\in[0,1]$.
%1. Gain to other firm is higher than own loss.
%    a. also implies consumer gains (?)
%2. Indirect effect: leads to higher uniform price by competitor, allowing firm to extract more from current customers.
%3. A priori, these reasons are not distinct, but in Hotelling model I think they are. This is because in second case, the consumer whose data is shared is harmed. (As are other consumers, by indirect effect.)

Firm $B$ can exploit both the indirect and direct effects of data sharing by sharing data both about consumers in $[\ul\theta, \mu(p_A)]$ and about consumers in $(\mu(p_A),1/2]$. Note, however, that sharing data about consumers in $(1/2, 1]$ is never beneficial, since it only results in a net loss to firms and has no indirect effect (by Lemma~\ref{lem:direct-effects}, above).

%\begin{lemma}\label{lem:baseprice}
%For every mechanism and any uniform price $p_A\geq 0$ of firm $A$, in every equilibrium consumers $\theta\in(1/2, 1]$ purchase from firm $B$.
%\end{lemma}
%\begin{proof}
%Fix a consumer $\theta\in(1/2,1]$. If both $A$ and $B$ offer $\theta$ a personalized price, then since $\theta$ is closer
%to $B$ the latter will always be able to offer a lower price. Thus, $\theta$ will purchase from $B$. If $A$ does
%not offer a personalized price, then consumer $\theta$ chooses between buying from $A$ at uniform price $p_A$ and getting utility
%$v-p_A - t\theta$, or buying from $B$ at personalized price $p_B(\theta)$ and getting utility 
%$v-p_B(\theta)-t(1-\theta)$. Since $\theta> 1/2$, firm
%$B$ can always choose a positive $p_B(\theta)$ such that $v-p_B(\theta)-t(1-\theta)> v-p_A - t\theta$.
%\end{proof}

%In this subsection, we proceed as follows. In Proposition~\ref{prop:firm-optimal} we describe the data-sharing mechanism and equilibrium that yield the firms maximal joint profits. We then show that, under this mechanism, there are other equilibria that are not as profitable. Finally, we consider variants of this optimal mechanism that do deliver high profits to both firms in all equilibria.
In general, the firm-optimal mechanism may depend on the distribution of consumers and other primitives of the model. In Proposition~\ref{prop:firm-optimal}, however, we show that when $v$ (the consumers' value for the good) is sufficiently high, then there is an essentially unique mechanism, with a corresponding equilibrium, that yield the firms maximal joint profits. The mechanism makes extreme use of the indirect effect of data sharing: Firm $B$ shares data about consumers $[0, 1/2]$, implying that firm $A$'s uniform price no longer applies to these consumers, and hence that this price can be almost arbitrarily high. $A$'s uniform price does apply to consumers in $(1/2, 1]$, for whom it serves as an outside option. However, because these consumers will always purchase from $B$ in equilibrium (by Lemma~\ref{lem:direct-effects}, above), the high outside option allows that firm to extract these consumers' entire surplus.

\begin{proposition}\label{prop:firm-optimal}
Fix $v>\frac{5t}{2(1-F(1/2))}$. Mechanism $M=([0,1/2],0)$ with equilibrium uniform price $p_A=v-t/2$ maximizes joint firm profits and is IR relative to any no-sharing equilibrium. Every other firm-optimal mechanism is of the form $M'=([0,1/2], r)$.
\end{proposition}

\begin{example}\label{ex:firm-optimal}
When consumers are uniformly distributed, the mechanism described in Proposition~\ref{prop:firm-optimal} is actually firm-optimal for all $v>2t$, as we now show.
This mechanism leads to profits $\pi_A=t/4$ and 
$$\pi_B=\int_{1/2}^1 (v-t(1-\theta))d\theta = \frac{v}{2} -\frac{t}{8}>\frac{7t}{8},$$
where the inequality follows since $v>2t$. Thus,  total profits are at least $9t/8$.
In contrast, consider any mechanism $M'=(M_B,r)$ in which firm $A$'s uniform price applies to a consumer in $[0,1/2]$, and fix some uniform price $p_A'\leq t$. Consumers $\theta\in (1/2,1]$ buy from $B$, leading to total profits at most $3t/4$ from these consumers. Consumers $\theta\in[0,1/2]$ either buy from $A$ at uniform price $p_A'$ or at personalized price $t(1-2\theta)$, or from $B$ at personalized price $t(2\theta-1)$ or $p_A'+t(2\theta-1)$ (depending on whether $\theta\in M_B$). Total profits are maximized when $p_A'=t$, consumers $\theta\in[0,1/4]$ buy at $A$'s personalized price, and the rest buy from $B$ at price $t+t(2\theta-1)$. Profits to $A$ from $[0,1/4]$ and to $B$ from $(1/4,1]$ are each equal to $3t/16$. Total profits from $M'$ are thus bounded above by $3t/4 + 2(3t/16) = 9t/8$, which is equal to the lower bound on profits from $([0,1/2],0)$.
\end{example}

% HAD RESULT WITHOUT BOUND ON F

%\begin{remark}
%Proposition~\ref{prop:firm-optimal-2} in Appendix~\ref{apx:firm-optimal} provides a general necessary and sufficient condition under which mechanism $M=([0,1/2],r)$ is IR for some $r$.
%\end{remark}

\begin{remark}
Proposition~\ref{prop:firm-optimal} provides a sufficient condition under which mechanism $M=([0,1/2],0)$ is firm-optimal for {\em some} equilibrium (namely, the one with uniform price $p_A=v-t/2$). However, under this mechanism there are other equilibria, which involve lower uniform prices, and that yield lower firm profits. In Proposition~6 in the full version of this paper \cite{gradwohl2023selling} we describe a different mechanism with $r=0$ that, while not firm-optimal, yields both firms strictly higher profits than under no sharing in {\em every} equilibrium.
\end{remark}

\section{Pareto-Improving Data-Sharing}\label{sec:pareto-improving}
In Section~\ref{sec:firm-optimal} above we show that firm $B$ can sell data in a way that maximizes joint firm profits, and hence allows that firm to charge a high price for the data. Such sharing, however, comes at the expense of consumers. In particular, under the equilibrium of Proposition~\ref{prop:firm-optimal}, firms extract the entire surplus of consumers located in $[1/2,1]$. In this section we show that there exist other data-sharing mechanisms that increase firm profits relative to no sharing, while at the same time also increasing consumers' utilities.

Recall that $P_A = \arg\max_{p}p\cdot F\left(\mu(p_A)\right)$, and that, by Proposition~\ref{prop:no-sharing}, the set of equilibria under no data-sharing consist of uniform prices $p_A\in P_A$ by firm $A$ and respective personalized prices $p_B(\theta)=\max\{0,  p_A+t(2\theta-1)\}$ by firm $B$. Denote by $E(p_A)$ the no-sharing equilibrium with uniform price $p_A$. In the following proposition we show that for each such no-sharing equilibrium there exists a Pareto-improving mechanism.

\begin{proposition}\label{prop:pareto-improving}
For every $p_A\in P_A$ there exists $r$ such that mechanism $M=\left(\left[\mu(p_A),\frac{1}{4}+\frac{\mu(p_A)}{2}\right],r\right)$ with uniform price $p_A$ is IR and weakly beneficial to every consumer, relative to $E(p_A)$. Furthermore, $M$ yields higher total firm profits than any other mechanism that is weakly beneficial to every consumer relative to $E(p_A)$.
\end{proposition}

The mechanism $M$ described in Proposition~\ref{prop:pareto-improving} does not decrease the utility of any consumer. Moreover, by bullet 2 of Lemma~\ref{lem:direct-effects}, that mechanism  {\em strictly increases} the utilities of a subset of consumers---namely, those located in 
$\left(\mu(p_A),\frac{1}{4}+\frac{\mu(p_A)}{2}\right]$.

The main idea underlying the construction for Proposition~\ref{prop:pareto-improving} is that firm $B$  shares data about every consumer $\theta$ that satisfies two conditions: (i) with no sharing, consumer $\theta$ prefers to pay $B$'s personalized price than $A$'s uniform price; (ii) sharing consumer $\theta$'s location leads to a net increase in firm profits. Note that these consumers are all closer to $A$ than to $B$, so that the welfare and profit gain is obtained due to an increase in efficiency. Finally, the construction is such that $A$'s uniform price under $M$ remains the same as with no sharing, which guarantees that consumers close to $A$ do not pay a higher price than under no sharing, but also that firms maximize their joint profits subject to this constraint.

\section{Consumer Opt-In}\label{sec:opt-in}
Proposition~\ref{prop:pareto-improving} above shows that there exist mechanisms that are strictly Pareto-improving, increasing firm profits as well as consumer welfare. However, these mechanisms are not optimal for firms---Proposition~\ref{prop:firm-optimal} identifies a different mechanism as maximizing firm profits, a mechanism that does so at the expense of consumers. How can a policymaker induce firms to share data in a Pareto-improving manner, rather than in a profit-maximizing manner? In this section we identify one way in which a policymaker can do this: by asking each consumer whether or not they agree to have their data shared, and then permitting firms to share data only about consumers who have agreed.

In order to analyze such consumer opt-in regulation, we first extend the model to include a preliminary opt-in stage. After setting up the model, we present two results. The first, Proposition~\ref{prop:opt-in} in Section~\ref{sec:opt-in-pareto}, states that, under consumer opt-in, there is an equilibrium of the extended model wherein firms choose the Pareto-improving mechanism of Proposition~\ref{prop:pareto-improving}. %The equilibrium is such that a certain segment of consumers refuses to opt in, and that, subject to this constraint, the profit-maximizing mechanism for the firms is the Pareto-improving mechanism.
Now, although consumer opt-in can lead to the choice of the Pareto-improving mechanism, there are other equilibria that do not. However, in our second result here---Proposition~\ref{prop:consumer-optimal} in Section~\ref{sec:opt-in-consumer-optimal}---we show that the equilibrium of Section~\ref{sec:opt-in-pareto}, where the Pareto-improving mechanism is chosen, is, in a sense, optimal for the consumers. 
%Thus, although consumer opt-in may lead to a multitude of equilibria, the one that is perhaps focal for consumers is the one that leads to our Pareto-improving mechanism.

\subsection{The Extended Model}\label{sec:opt-in-model}
We begin by extending the model of Section~\ref{sec:model} with a preliminary stage, in which each consumer simultaneously chooses whether or not to opt in to having location data shared. Denote the set of consumers who opted in as $C$. Only then do firms engage in a data-sharing mechanism $M=(M_B,r)$; however, firms are restricted to choosing a mechanism for which $M_B\subseteq C$. Such mechanisms are {\em feasible for $C$}.

We assume that firms bargain over the choice of mechanism efficiently---that is, they choose a mechanism $M$ that maximizes total firm profits, subject to the opt-in constraint. One way to implement such efficient bargaining is when one of the firms makes the other a take-it-or-leave-it offer by suggesting a mechanism $(M_B, r)$ that is feasible for $C$. Depending on which firm makes the offer, the chosen price transfer $r$ will vary to favor the offering firm. Either way, however, firms will choose to offer a mechanism that maximizes joint firm-profits. This assumption is stated formally in Definition~\ref{def:tfne} below as part of the solution concept. In addition, as in the previous sections, we assume that, absent data-sharing, firms play the no-sharing equilibrium $E(p_A)$ for some $p_A\in P_A$. 

In this extended model there is an additional, technical complication. We are assuming that firms choose a mechanism that is feasible for some $C$. However, since $C$ is generated by the set of consumers who choose to opt in, it may not be a measurable set. Thus, the firms' optimization problem may not be well-defined at every $C$. One way to get around this problem is to consider the Nash equilibria of this extensive-form game (rather than the subgame perfect equilibria). However, this is somewhat unsatisfying, as such equilibria may be sustained by strange off-equilibrium behavior---namely, the presence of empty threats. Instead, we will use an equilibrium notion that is weaker than subgame perfect equilibrium but nonetheless suffices to eliminate empty threats. The general definition is due to \cite{gradwohl2013sequential}; here we give a specialized version that applies to our specific game.

For the definition, let $L$ denote the set of %Lebesgue measurable 
subsets of $[0,1]$, and let $\scM(C)$ denote the set of all mechanisms feasible for $C$.
\begin{definition}\label{def:tfne}
A set $C^*\in L$ and functions $m:L\rightarrow \scM(L)$ and  $p:L\rightarrow \real_+$ form a {\em threat-free Nash equilibrium (TFNE)} if
\begin{enumerate}
\item For every $C$, mechanism $m(C)$ is feasible for $C$, and $p(C)$ is an equilibrium uniform price for firm $A$ under $m(C)$.
\item For every $\theta\in C^*$, consumer $\theta$ is weakly better off under $m(C^*)$ than under $m(C^*\setminus\{\theta\})$ (with respective uniform prices $p(C^*)$ and $p(C^*\setminus\{\theta\})$).
\item For every $\theta\in[0,1]\setminus C^*$, consumer $\theta$ is weakly better under $m(C^*)$ than under $m(C^*\cup \{\theta\})$  (with respective uniform prices $p(C^*)$ and $p(C^*\cup\{\theta\})$).
%\item $m(C^*)$ is IR and jointly firm-optimal relative to all mechanisms that are feasible for $C^*$.
\item For every $\theta\in [0,1]$ and $C\in\left\{C^*\setminus\{\theta\},C^*\cup\{\theta\}\right\}$, mechanism $m(C)$ with uniform price $p(C)$ is IR and jointly firm-optimal relative to all mechanisms that are feasible for $C$ (with corresponding uniform prices).
\end{enumerate}
\end{definition}

For comparison, in a Nash equilibrium bullet 4 would be replaced by requiring IR and joint firm-optimality only for $C^*$. In a subgame perfect equilibrium, in contrast, bullet 4 would require these for all sets $C$. A TFNE is a compromise between the two, requiring IR and joint firm-optimality for $C^*$ and for all sets $C$ that differ from $C^*$ by a single consumer's unilateral deviation. 

\subsection{Pareto-Improving Equilibrium}\label{sec:opt-in-pareto}
Given the extended model above, we can now state our  proposition on the benefit of consumer opt-in.

\begin{proposition}\label{prop:opt-in}
For every $p_A\in P_A$ there exists a TFNE $(C^*,m,p)$ of the extended model in which $m(C^*)$ is the Pareto-improving mechanism $M=\left(\left[\mu(p_A),\frac{1}{4}+\frac{\mu(p_A)}{2}\right],r\right)$, for some $r$.
\end{proposition}

Proposition~\ref{prop:opt-in} shows that, when consumers can choose whether or not to opt in to having their data shared, and firms are allowed to only share the data of consumers who have opted in, then the equilibrium mechanism is Pareto improving. There are other equilibria that lead to a Pareto-improving mechanism. In fact, as long as consumers 
$\theta\in[0,\mu(p_A))$ do {\em not} opt in to having their data shared, the mechanism that maximizes firms' profits will be Pareto improving. 

However, there are also other equilibria in which the chosen mechanism is not Pareto improving.  Consider the following strategies: Consumers $[0,1/2]$ opt in to having their data shared, and firms choose the firm-optimal mechanism $M=([0,1/2], 0)$ from Proposition~\ref{prop:firm-optimal}. If some consumer $\theta\in[0,1/2]$ does not opt in, then firms use the mechanism $M_\theta=([0,1/2]\setminus\{\theta\}, 0)$. This mechanism is identical to $M$, except that consumer $\theta$ faces firm $A$'s uniform price $p_A=v-t/2$ rather than the personalized price $p_A(\theta)=t(1-2\theta)$. This is no better for consumer $\theta$, and so these strategies form an equilibrium. Why, then, would consumers choose to collectively opt in as in Proposition~\ref{prop:opt-in}? 

\subsection{Consumer-Optimal Equilibrium}\label{sec:opt-in-consumer-optimal}

We now show that the equilibrium of Proposition~\ref{prop:opt-in} is focal for the consumers. In particular, we show that it maximizes consumer welfare, relative to all other equilibria that leave no consumer worse off.
%For simplicity, suppose that the set of equilibrium uniform prices $P_A$ is a singleton, $P_A=\{p_A\}$.  

Fix some $p_A\in P_A$, and observe that there always exists a TFNE of the extended game in which no consumer opts in, and that this leads to consumer utilities as derived from equilibrium $E(p_A)$ in mechanism $M_\emptyset=(\emptyset, 0)$.
Next, let us  consider other opt-in choices for consumers. For a set $C$ and mechanism $M$ feasible for $C$, say that $M$ is {\em Pareto-improving for the consumers} if the resulting utility of every consumer is weakly higher than under $M_\emptyset$.
We now show that consumers' utilities in the equilibrium of Proposition~\ref{prop:opt-in} are optimal:

\begin{proposition}\label{prop:consumer-optimal}
Fix $C\subseteq[0,1]$ and a mechanism $M$ with uniform price $q_A$ that is feasible for $C$ and that is Pareto-improving for the consumers. If $M$ yields strictly higher total utility to the consumers than $M^*$, then $M$ will not be chosen by the firms in any TFNE in which consumers $C$ opt in.
\end{proposition}

That is, if we assume consumers make their opt-in decisions in a way that leads to a weak improvement for each, then they can do no better than the opt-in strategy of Proposition~\ref{prop:opt-in}.

\section{Conclusion}\label{sec:conclusion}
In this paper we analyzed the benefits to a data-holder of selling consumer data to a data-buyer in a Hotelling  model of imperfect competition. We identified the two effects of data sharing, and showed that the interplay of these effects can lead to Pareto-improving mechanisms that benefit consumers as well as firms. Finally, we showed that consumer opt-in can induce firms to choose such a Pareto-improving mechanism.

\section*{Acknowledgements} The work of Tennenholtz was supported by funding from the European
Research Council (ERC) under the European Union's Horizon 2020
research and innovation programme (grant number 740435).

\bibliographystyle{eptcs}\bibliography{hotellingDS}
\end{document}